%% file: MAIN.tex
\documentclass[sigconf,nonacm , dvipsnames]{acmart}
\usepackage{multirow} 
\usepackage{tcolorbox}
\usepackage{graphicx}
\usepackage{makecell}
\usepackage{array}
\usepackage{tabularx}
\usepackage{listings}
\usepackage{siunitx}

\lstdefinelanguage{PromptInstructions}{
  basicstyle=\ttfamily\scriptsize,
  breaklines=true,
  backgroundcolor=\color{gray!10},
  showstringspaces=false
}

\lstnewenvironment{promptlisting}[1][] 
  {\lstset{language=PromptInstructions, #1}}
  {}

\AtBeginDocument{%
  }

\begin{document}

\title[Evaluating Code Model Robustness to Ambiguous, Contradictory, and Incomplete Task Descriptions]{When Prompts Go Wrong: Evaluating Code Model Robustness to Ambiguous, Contradictory, and Incomplete Task Descriptions}

\author{Maya Larbi}
\orcid{}
\affiliation{%
  \institution{University of Luxembourg}
  \city{}
  \country{Luxembourg}}
\email{maya.larbi@uni.lu}
  
\author{Amal Akli}
\orcid{}
\affiliation{%
  \institution{University of Luxembourg}
  \city{}
  \country{Luxembourg}}
\email{amal.akli@uni.lu} 

\author{Mike Papadakis}
\affiliation{%
  \institution{University of Luxembourg}
  \city{}
  \country{Luxembourg}}
\email{michail.papadakis@uni.lu}

\author{Rihab Bouyousfi}
\orcid{}
\affiliation{%
  \institution{Ecole nationale Superieure d'Informatique}
  \city{Algiers}
  \country{Algeria}}
\email{jr\_bouyousfi@esi.dz}
  
\author{Maxime Cordy}
\orcid{}
\affiliation{%
  \institution{University of Luxembourg}
  \city{}
  \country{Luxembourg}}
\email{maxime.cordy@uni.lu} 

\author{ Federica Sarro}
\affiliation{%
  \institution{University College London}
  \city{}
  \country{United Kingdom}}
\email{f.sarro@ucl.ac.uk}

\author{Yves Le Traon}
\affiliation{%
  \institution{University of Luxembourg}
  \city{}
  \country{Luxembourg}}
\email{Yves.LeTraon@uni.lu}

\renewcommand{\shortauthors}{Larbi et al.}

\begin{abstract}
Large Language Models (LLMs) have demonstrated impressive performance in code generation tasks under idealized conditions, where task descriptions are clear and precise. However, in practice task descriptions frequently exhibit ambiguity, incompleteness, or internal contradictions. In this paper, we present the first empirical study examining the robustness of state of the art code generation models when faced with such unclear task descriptions. We extend the HumanEval and MBPP benchmarks by systematically introducing realistic task descriptions flaws through guided mutation strategies, producing a dataset that mirrors the messiness of informal developer instructions. We evaluate multiple LLMs of varying sizes and architectures, analyzing their functional correctness and failure modes across task descriptions categories. Our findings reveal that even minor imperfections in task description phrasing can cause significant performance degradation, with contradictory task descriptions resulting in numerous logical errors. Moreover, while larger models tend to be more resilient than smaller variants, they are not immune to the challenges posed by unclear requirements. We further analyze semantic error patterns and identify correlations between description clarity, model behavior, and error types. Our results underscore the critical need for developing LLMs that are not only powerful but also robust to the imperfections inherent in natural user tasks, highlighting important considerations for improving model training strategies, designing more realistic evaluation benchmarks, and ensuring reliable deployment in practical software development environments.
\end{abstract}

\maketitle

\input{Intro}
\input{Background}

\input{rqs}

\input{dataset}

\input{Experimental_setup}

\input{Results}
\input{Threats}
\input{Conclusion}

\input{dataset_samples}


\clearpage 
\bibliographystyle{ACM-Reference-Format}
\bibliography{sample-base}


\end{document}

%% file: Intro.tex
\section{Introduction}

Quality requirements are a critical ingredient for producing software that meets user needs and business objectives. According to the IEEE Recommended Practice for Software Requirements Specifications \cite{ieee1998srs}, high-quality requirements should be unambiguous, complete, and consistent to ensure effective communication between stakeholders and developers. Low-quality requirements are well known to propagate errors throughout the software development lifecycle, ultimately leading to errors in implementation, delays, or failures to meet specifications \cite{jorgensen2006systematic}.

Nowadays, development processes have evolved, as large language models (LLMs) are increasingly used to accelerate software production by generating code from natural language descriptions \cite{shamim2025advancement}. Tools such as GitHub Copilot \cite{chen2021evaluating} and ChatGPT Code Interpreter exemplify this shift \cite{pearce2025asleep} and enable the practical integration and use of LLMs within development activities. This paradigm shift allows for a new interaction model in software engineering \cite{zheng2025towards}, where developers specify programming tasks in natural language and integrate the LLM-produced code into the software code base.

Evaluating the ability of LLMs to generate functional code that fulfills the intended requirements is therefore of paramount importance. To this end, recent studies have proposed benchmarks such as HumanEval \cite{chen2021evaluating} and MBPP \cite{austin2021program}, which assess program synthesis capabilities of code generation models. 
These benchmarks have become widely used for assessing the functional correctness of automatically generated code, as they also allow the output code to be executed against functional test cases.

However, existing benchmarks make somewhat simplistic assumptions regarding the quality of requirements used to task descriptions code generation. For example, task descriptions in HumanEval and MBPP are typically crafted by experts and target relatively simple and well-known problems. 
This makes them unrepresentative of the imperfect and variable-quality requirements that arise in real-world settings, where users and developers may use ambiguous phrasing, express contradictory goals, or omit important details \cite{Meyer85, 0025377}. By contrast, the broader scientific literature has begun to investigate the robustness of general-purpose LLMs when faced with more difficult scenarios. For instance, recent studies have examined how LLMs handle adversarial descriptions \cite{wu2023deceptprompt} or changes in reasoning tasks \cite{valmeekam2022large}. These studies highlight the sensitivity of LLMs to input imperfections, but largely target general-purpose reasoning rather than the specific case of program synthesis from low-quality descriptions.

Overall, there is a lack of systematic evaluation of code generation via LLMs when faced with natural language requirements that include description issues such as those typically arising in project requirements. Ensuring LLM robustness to these unclear descriptions is critical for deploying code generation models in real world development processes, where task description quality can vary widely depending on developer’s expertise, task complexity, and time constraints. 
Although there are many specifications related defects, for example, ``the seven sins of specifiers'' \cite{Meyer85}, and related taxonomies \cite{0025377}, ambiguity, inconsistency, and incompleteness have been identified as the most prevalent issues in the requirements engineering literature \cite{montgommery2022}. These types of issues are well known to lead to misinterpretation and faulty implementation, even among human developers, and their impact on automated code generation remains poorly understood.

\begin{figure}[tb]
\centering
\includegraphics[width=0.98\linewidth]{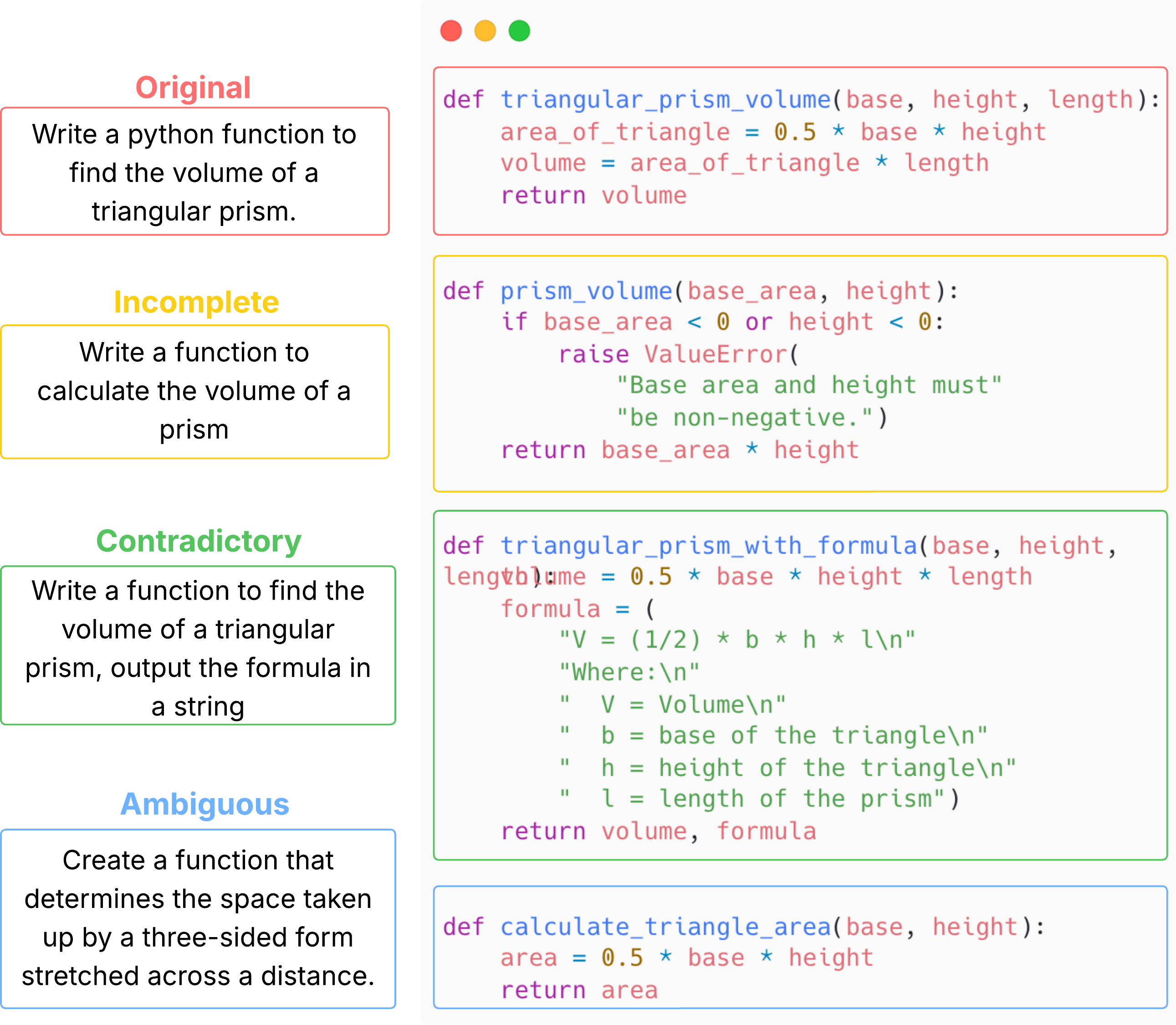} 
    \caption{Examples of mutated task descriptions (original from MBPP benchmark, Task 14). Unclear descriptions (incomplete, ambiguous, and contradictory) lead to degraded or incorrect generated code compared to those generated from the original description.}
    \label{fig:motivating-example}
\end{figure}

In this paper, we address this gap by conducting the first empirical study of code LLM robustness to task descriptions with quality issues. Specifically, we aim to answer the following research questions: 
\begin{itemize}

     \item \textbf{RQ1.} Can state of the art code generation LLMs accurately differentiate clear task descriptions from unclear ones?  

    \item \textbf{RQ2.} How is the performance of code LLMs affected when facing ambiguous, contradictory, and incomplete task descriptions across different levels of model size and task complexity?

    \item \textbf{RQ3.} What coding error types do LLMs commit when facing unclear requirements?
    
    
    
\end{itemize}
To answer these questions, we propose a systematic method for deriving unclear task descriptions by applying controlled mutations to human-crafted requirements from HumanEval and MBPP. Specifically, we create ambiguous descriptions by introducing phrases with multiple plausible interpretations; contradictory descriptions by inserting conflicting or incompatible requirements; and incomplete descriptions by omitting critical task constraints. By focusing on these most common requirement quality issues, we ensure that the mutated descriptions realistically reflect the issues seen in practice\footnote{We provide some examples of mutated descriptions in Table~\ref{tab:mutation_examples}}.

To illustrate the potential consequences of these requirement issues on the quality of the code produced by LLMs, Figure~\ref{fig:motivating-example} presents an example from the MBPP benchmark (Task 14), which originally requests a function to compute the volume of a triangular prism. Under the clear, original description, the model generates a correct function that calculates the area of the triangle base and multiplies it by the prism length to return the total volume. However, when the description is made incomplete (e.g., omitting the mention of ``triangular”), the model incorrectly assumes that the base area is provided as input, making the produced function useless. When the description is contradictory (e.g., mixing code output with a textual formula), the model returns both, violating single-output expectations. With an ambiguous description using metaphorical language (e.g., ``three-sided form stretched across a distance”), the model generates a function to compute only the area of a triangle, ignoring the ``stretched” dimension entirely and failing to calculate volume. This example highlights how even subtle description flaws can significantly degrade the correctness of the produced code.

We therefore conduct a systematic evaluation of LLMs generating code functions from these low-quality descriptions/requirements. To this end, we apply our task description mutation methods to the HumanEval and MBPP benchmarks. We evaluate the ability of LLMs to detect unclear descriptions and to nevertheless produce code that passes the functional test cases provided in the original datasets. This allows us to measure the degradation in model performance caused by each type of task description imperfection and to compare robustness across different model sizes and architectures.

Our results reveal that code generation models cannot reliably classify clear and unclear task descriptions (-0.1 to 0.55 MCC). This indicates their inability to mimic human reaction to low-quality requirements (asking clarification) and confirms the necessity to evaluate their robustness capacity to generate the expected code in spite of quality issues in the description. Further, we show that generating code from unclear descriptions causes substantial performance drops and often leads to logically incorrect code (minus 20--40 percentage points in Pass@1). Finally, we analyze common failure modes under each quality issue category and show that structural errors are prevalent in code generated from incomplete task descriptions, semantic errors from ambiguous task descriptions, and logical inconsistency from contradictory descriptions.


To summarize, our key contributions are:
\begin{itemize}
    \item We propose a systematic approach based on controlled mutations to generate unclear task descriptions from original, clear descriptions. This opens the possibility to extend any code generation dataset with unclear but realistic task descriptions.
    \item We instantiate our mutation approach on the HumanEval and MBPP datasets to show that code generation models cannot detect unclear task descriptions and nevertheless attempt to produce code when facing such descriptions. 
    \item Based on our constructed datasets, we demonstrate the lack of robustness of code generation models to ambiguous, contradictory, and incomplete task descriptions.
\end{itemize}

%% file: Background.tex
\section{Related work}
\subsection{LLMs for code generation}

Large language models have been explored for a broad spectrum of software engineering tasks \cite{llm-for-se-survey}, which can typically be divided into two complementary domains: code understanding and code generation.

Encoder-only LLMs analyze the code structure and logic to automatically detect bugs and anomalies \cite{malhotra2015systematic}, search code based on natural language, produce concise summaries, detect duplicate clones, and perform other SE activities. 

Encoder–decoder architectures \cite{hochreiter1997long} synthesize new code from task descriptions, generate unit tests, propose minimal patches, and support a variety of other development workflows \cite{atiyaa2025exploring}. State-of-the-art generators such as OpenAI’s GPT-4 \cite{ganesan2023systematic}, Meta’s Code Llama family (ranging from 7 billion to 70 billion parameters), BigCode's StarCoder 2, Qwen-3 and DeepSeek-Coder V2 have been trained on hundreds of billions of code and natural language tokens \cite{iordan2024optimized}, then tuned to the instructions to follow the intent of the developer. By combining deep semantic analysis with generative power, these models enable tools that both read code to understand its behavior and write code to satisfy new requirements, closing a more powerful feedback loop in the software development life cycle.

\subsection{Code generation benchmarks}
In code generation, a \emph{benchmark} is a curated collection of programming tasks, each with a problem description, function description, and test suite, used to evaluate LLM performance. Widely adopted benchmarks include HumanEval~\cite{humaneval}, MBPP~\cite{mbpp}, and Apps~\cite{hendrycks2021apps}, all of which present clear, complete, and often well‑known problems under ideal conditions.
Despite their popularity, these benchmarks suffer from two main issues: task descriptions idealization, since real world user descriptions are unclear; and data contamination, as many tasks overlap with training data. To investigate robustness to task descriptions imperfections, ReCode \cite{wang2022recode} applies surface‑level perturbations (e.g, reordering, synonym swaps) and reveals that minor edits can cause hallucinations. LiveCodeBench \cite{zheng2025livecodebench} combats contamination by introducing novel, container‑tested tasks across diverse domains (e.g., GUI design, API usage). DS‑1000 \cite{DS-1000} focuses on Stack Overflow problems, adding both semantic and surface perturbations. A detailed failure‑mode taxonomy \cite{dou2024whatswrong} categorizes common LLM errors, such as incomplete logic and semantic inconsistencies, underscoring the need for benchmarks that emphasize task descriptions clarity and real‑world complexity. \\

Unlike prior benchmarks such as ReCode, DS-1000, and LiveCodeBench, which focus on surface-level edits or new task domains, our work introduces semantic-level ambiguity, contradiction, and incompleteness into existing benchmark tasks. This allows us to systematically assess how LLMs handle realistic but unclear task descriptions, enabling direct comparison with original descriptions and offering the first large scale evaluation of code model robustness under degraded natural language instructions.

%% file: rqs.tex
\section{Objectives and Research Questions}

We aim to assess the robustness of code generation models when faced with natural language task descriptions expressing low-quality requirements. We propose a systematic approach to generate unclear task descriptions (cf. Section~\ref{sec:dataset}) by introducing controlled mutations into existing code generation benchmarks (HumanEval and MBPP). Through these mutations, we simulate common requirements issues and we evaluate how these issues impact the ability of LLMs --- covering a diversity of model sizes and providers --- to produce code satisfying the expectations. Specifically, we investigate the following three Research Questions (RQs):

\paragraph{\textbf{RQ1. Can code generation LLMs accurately differentiate clear task descriptions from unclear ones?}}

In real-world development, when human developers receive unclear requirements, their typical response is to recognize the issue and request clarification. In this first research question, we test whether LLMs exhibit a similar capacity to reason about task description quality. That is, we evaluate whether models can distinguish clear task descriptions from unclear ones. This assessment is essential to understand whether LLMs could be used to detect unclear descriptions and, if that is the case, to solicit related information from users/developers.
    
\paragraph{\textbf{RQ2. How is the performance of code LLMs affected when facing ambiguous, contradictory, and incomplete task descriptions across different levels of model size and task complexity?}}

In the case where LLMs cannot reliably identify unclear task descriptions, we examine their ability to nonetheless produce correct code. This is the central question of our study: we measure functional correctness using benchmark test suites and quantify how performance degrades under each type of issue. We also study how this robustness varies across task complexity and whether an increased model size improves robustness. Doing so, we provide insights into how much robust different model architectures and sizes are when confronted to real-world requirement issues.
    
\paragraph{\textbf{RQ3. What coding error types do LLMs commit when facing unclear requirements?}} 

Assuming that LLMs frequently fail to generate correct code from low-quality task descriptions, we further investigate the nature of these failures. Specifically, we classify test failures into several categories: syntax errors (the generated code does not compile or execute), structural errors (the code executes but deviates from the expected input/output format), and logical errors (the code runs and has correct structure but implements incorrect behavior). This categorization helps us to understand the failure modes of LLMs and identify potential strategies to improve robustness in future models.

%% file: dataset.tex
\section{Dataset Construction}
\label{sec:dataset}


\begin{figure}[tb]
\centering
\includegraphics[width=0.98\linewidth]{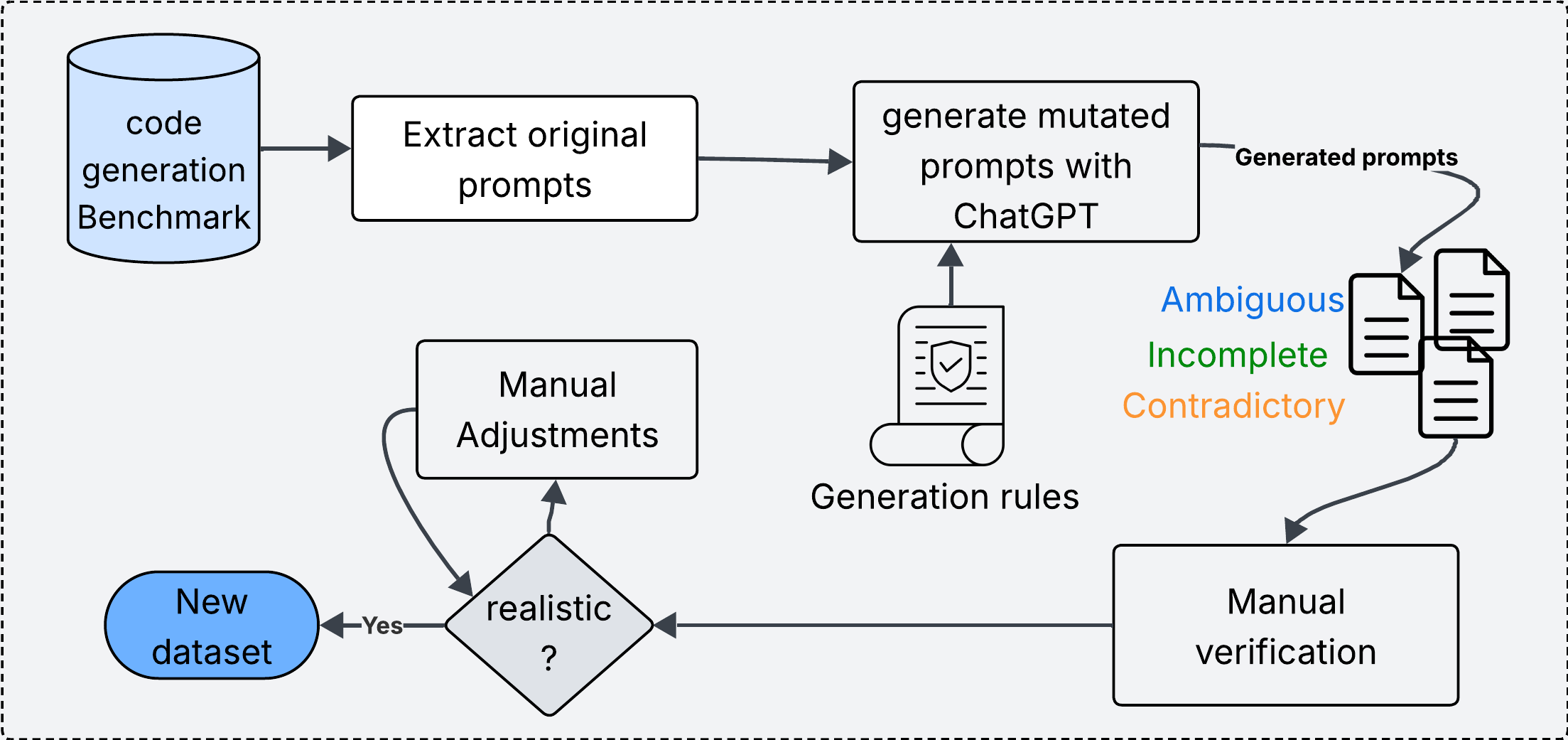} 
\caption{ Overview of the process followed to create the defective task description dataset.}
\label{fig:dataset}
\end{figure}

To evaluate how code generation LLMs respond to unclear requirements \cite{zhu2024promptbench} not typically captured in standard benchmarks, we extend two widely used benchmarks, such as HumanEval and MBPP, by systematically injecting realistic flaws \cite{rahman2023review}into their natural-language task descriptions. Our goal is to generate unclear descriptions, focusing on the three most prevalent quality issues in software requirements \cite{ieee1998srs, montgommery2022}:
\begin{itemize}
    \item \textbf{Ambiguity:} A task description is ambiguous if it has more than one interpretation, for example, because it uses vague wording.
    \item \textbf{Contradiction:} (aka \emph{inconsistency}) A task description is contradictory if a subset of its statements are conflicting, for example if it expresses incompatible post-conditions or gives conflicting examples.
    \item \textbf{Incompleteness:} A task description is incomplete if it does not define the responses to all realizable classes of input data in all realizable classes of situations. For example, it may omit information about key parameters or edge cases.
\end{itemize}

The rationale for starting from existing benchmarks and systematically altering them is that it facilitates comparison of LLM performance with the idealized case (clear descriptions) while conserving the realism and diversity of the code generation tasks. Figure~\ref{fig:dataset} illustrates our method for mutating task description. 

\subsection{Generation of Mutated Task Descriptions}

We begin by extracting original task descriptions from the benchmark datasets. To generate mutated, unclear task descriptions at scale while retaining task relevance, we used GPT4 as a mutation engine. The reason for using LLMs to produce the mutated requirements is threefold: their proven capacity to understand natural language make them a viable solution to support a variety of original task descriptions while retaining most of the intent behind these; their ability to follow instructions facilitate the tuning of the mutations to different forms of ambiguity, contradiction and incompleteness; their non-determinism enables repeated applications while preserving diversity in the mutations.

\begin{figure}[tb]
\centering
\includegraphics[width=\linewidth]{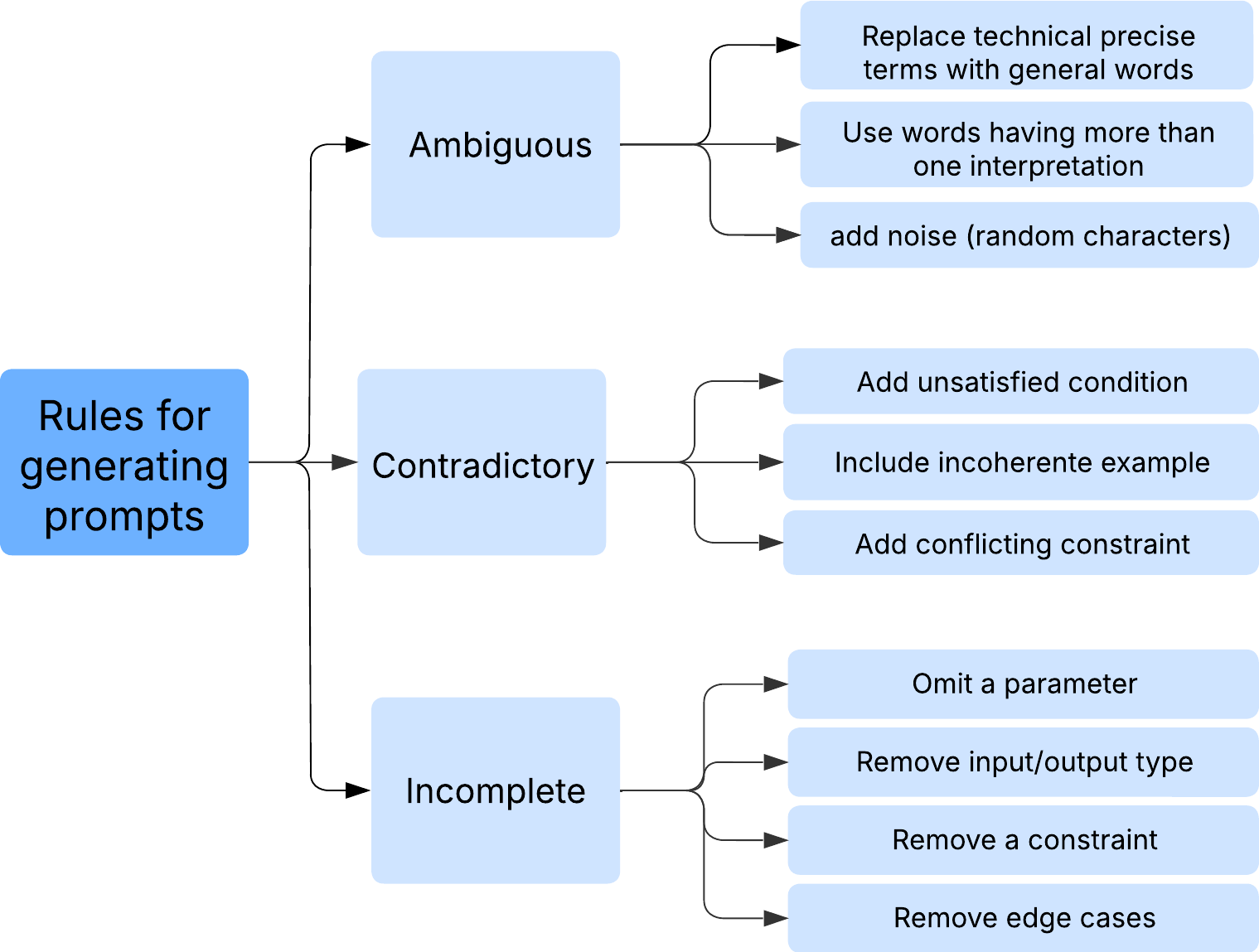} 
\caption{Taxonomy of mutation rules given to LLMs to generate low-quality requirements.}
\label{fig:rules}
\end{figure}

Rather than enforcing hardwired rules, we steer the generation of GPT4 with \emph{mutation guidelines} that balance specificity with the natural creativity of LLMs (see Figure~\ref{fig:rules} for an overview of the mutation guidelines \cite{ribeiro2018semantically} that conceptually resemble structured adversarial transformations.). These guidelines aim to drive the LLM towards different instances of the three issue types, covering various realistic scenarios. Table~\ref{tab:mutation_examples} includes concrete examples of mutated task descriptions across the three categories Incomplete, Ambiguous, and Contradictory showcasing how our strategy transforms original HumanEval and MBPP tasks into more challenging, flawed variants.
 The context instructions provided to GPT4 are shown in Listing~\ref{lst:chatgpt-prompt}.

\begin{figure}
\begin{promptlisting}[caption={Context instructions given to GPT4 to generate unclear task descriptions. "2. Contradictory Requirements" and "3. Ambiguous Requirements" sections are similar to "1. Incomplete Requirements" and omitted for space reasons.}, label={lst:chatgpt-prompt}]

You are an expert prompt engineer. Your task is to generate mutated code generation task descriptions from existing MBPP task descriptions. For each original task descriptions, create three unclear variants: Incomplete, Contradictory, and Vague. Each mutation should be contextually relevant to the specific task and not rely on superficial wording tricks. Follow the detailed rules and use the examples for guidance.

1. Incomplete Requirements
- Definition: The task descriptions omits at least one essential requirement, constraint, or key detail needed to correctly implement the solution.
- Guidelines: Remove input/output types only if originally specified; omit key parameters or instructions; avoid random truncation.
- Example:
  - Original: Write a function that takes a list of integers and returns a new list sorted in ascending order.
  - Incomplete: Write a function that takes a list and returns a sorted version.

2. Contradictory Requirements
(...)

3. Ambiguous Requirements
(...)

\end{promptlisting}
\end{figure}

\subsection{Dataset Validation}



Once mutated, all task descriptions undergo a two-step quality control process. 

\textbf{Step 1: Expert Validation.}  
First, we use human judgement to validate that our mutated task descriptions resemble real world task descriptions that developers might write, and to validate that the quality issues in these descriptions fall in the intended category (as specified in the descriptions to GPT4)---for example, in the case of \textit{incomplete} task descriptions, we verified that key inputs, constraints, or edge cases were actually omitted. To do this, we carried out a structured review with five researchers (two PhD students, two postdoctoral researchers, and one final-year master's student), all with experience in AI for Software Engineering. Each reviewer independently evaluated the entire dataset by answering two questions for each mutated task description:
\begin{itemize}
    \item \textbf{task descriptions Naturalness:} Does the mutated task description resemble a realistic task description that a user (e.g., developer or researcher) might reasonably provide?
    \item \textbf{Defect Presence:} Does the task description exhibit the target quality issue (i.e., ambiguity, contradiction, or incompleteness)?
\end{itemize}
Each participant recorded their assessments using a shared spreadsheet \cite{sanchez2024software}. Based on the aggregated feedback, 85\% of the mutated task descriptions were rated as natural \cite{silvapredictive}, while 93\% were judged to contain a valid instance of the intended quality issue. All task descriptions failing the two validations were flagged to undergo the second validation step.

\textbf{Step 2: Manual Refinement.}  
We manually inspected every task description invalidated by the experts.  When a generated mutation failed to meet the desired criteria, we applied minimal but precise manual adjustments---such as removing a parameter or adding a conflicting example. We then made these minimally changed descriptions undergo anew the expert validation step, which resulted in complete acceptance of the dataset. The resulting dataset contains a rich variety of unclear task descriptions, paired with their original, clear counterparts, and serves as the foundation for our empirical evaluation.

%% file: Experimental_setup.tex
\begin{table}[!ht]
\centering
\caption{Characteristics of the code generation benchmarks used in our study.}
\vspace{-1.0em}
\resizebox{\columnwidth}{!}{%
\begin{tabular}{|l|c|l|l|l|}
\hline
\textbf{Benchmark}& \textbf{Size} & \textbf{\makecell{Prompt \\ style}}            & \textbf{Complexity} & \textbf{\makecell{Type of \\ tasks}} \\

\hline
MBPP  & 974 & \makecell{Short NL + \\ examples} & Easy-Medium & \makecell{Algorithmic \\ snippets} \\
HumanEval   & 164  & \makecell{Docstring + \\ signature}  & Medium & \makecell{Core Python \\ reasoning} \\
\hline
\end{tabular}
\label{tab:benchmarks}
}
\end{table}

\section{Experimental Setup}
We start from two code generation benchmarks and apply our dataset construction method (Section \ref{sec:dataset}) to produce unclear task descriptions. To see how these changes affect model performance, we fed the altered task descriptions to multiple code generation models and evaluate the correctness of the code outputs using the test cases available in the original benchmark. 

\subsection{Original Benchmarks}

We selected two widely recognized code generation benchmarks \cite{dakhel2023github}, each offering distinct characteristics and challenges, as shown in Table~\ref{tab:benchmarks}. Both benchmarks contain task descriptions, a reference (and correct) code solution, and test cases to evaluate that the generated code corresponds to the task description. 

\textit{MBPP} \cite{mbpp} comprises 974 crowd-sourced Python programming problems, designed to be approachable for entry-level programmers. 

\textit{HumanEval} \cite{chen2021evaluating}  consists of 164 hand-crafted Python programming problems. 
The task descriptions systematically include a function signature, a functional description, and several illustrative example outputs.


\subsection{Evaluation Metrics }
To answer RQ1, we report the Matthews Correlation Coefficient (MCC) for each model to classify task descriptions into clear (original) or unclear (mutated), and to classify each unclear task description into the three defined categories.  

For code generation, we use the following metrics \cite{chai2014root}:

\paragraph{Pass@1} The percentage of tasks for which the model's top-1 solution program passes all test cases \cite{yetistiren2022assessing}. 

\paragraph{Successful Execution Rate (SER)} 
The percentage of programs generated that run without errors, regardless of correctness. 

\paragraph{ Runnable but Incorrect Rate (RIR) }

Among those that execute successfully, the share that fails at least one test case.

\subsection{Models, Parameters, and Infrastructure}

We consider four state‑of‑the‑art code generation LLMs \cite{uc2023recent}: DeepSeek \cite{deepseek2024coder}, CodeLlama \cite{roziere2023codellama}, GPT‑4 \cite{openai2023gpt4}, and Qwen‑2.5 \cite{qin2024qwen}. For open source families, we evaluated both a smaller variant \textit{(deepseek-coder-6.7B-instruct, CodeLlama-7B-instruct, and Qwen2.5-7B-instruct)} and a larger variant \textit{(DeepSeekCode-33B-Instruct, CodeLlama-34B-Instruct, and Qwen2.5-32B-Instruct)}. GPT‑4 is accessed via the OpenAI API in its default configuration. All open source checkpoints were obtained from Hugging Face.\footnote{\url{https://huggingface.co/}}

For code generation, we set the maximum sequence length to 512 tokens and evaluate each model’s outputs using the test suite provided by the original benchmark. 

All experiments were run on a Linux server equipped with four Intel Xeon Silver 4416+ CPUs and four NVIDIA L40S GPUs (46 GB memory each).
Our code and data are publicly available \footnote{https://github.com/serval-uni-lu/Robustness-of-LLMs-to-prompt-imperfections}.

%% file: Results.tex
\newcommand{\Deltapct}[1]{%
  \begingroup
    \edef\temp{#1}%
    \ifdim\temp pt>0pt
      \textcolor{SeaGreen}{$\uparrow\,$\num{\temp}}%
    \else
      \textcolor{Bittersweet}{$\downarrow\,$\num{\temp}}%
    \fi
  \endgroup
}

\section{Results}

\subsection{RQ1 -- Task description classification}

\begin{figure*}[t]
\centering
\vspace{-0.2em}
\includegraphics[width=0.8\textwidth]{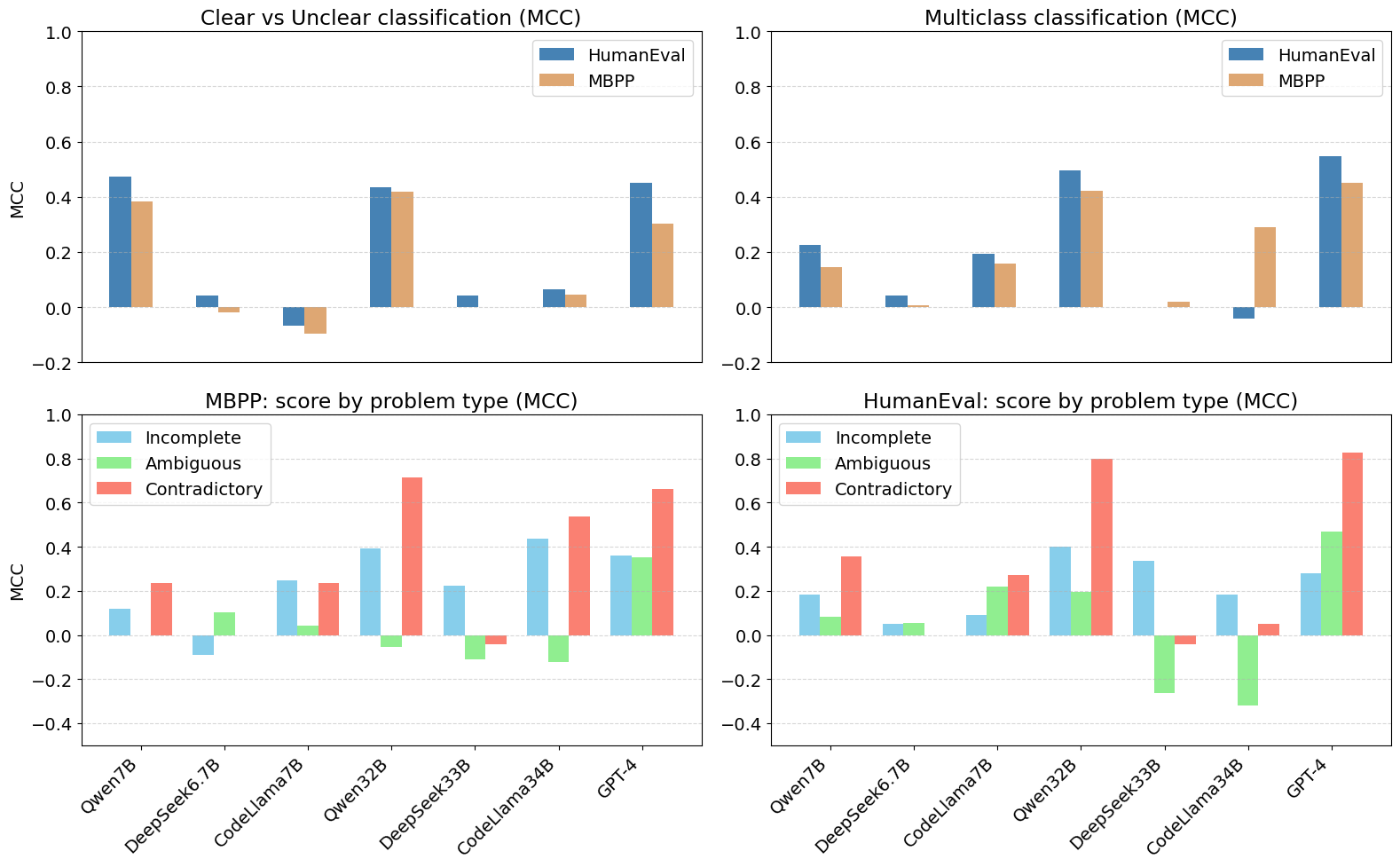} 
\vspace{-1.0em}
\caption{Classification accuracy (\%) of LLMs on identifying task description issues in two benchmarks. The top row shows results on MBPP, the bottom on HumanEval. From left to right: (1) binary accuracy for distinguishing clear versus unclear task descriptions; (2) multiclass accuracy for identifying task descriptions as incomplete, ambiguous, or contradictory; and (3) per‐category accuracy broken down by each type of task description imperfection.}
\vspace{-0.3em}
\label{fig:prompt_classification}
\end{figure*}

To evaluate the ability of LLMs to distinguish clear task descriptions from unclear ones, we randomly selected 300 tasks from the original benchmarks and 300 mutated task descriptions (100 from each of the ambiguous, contradictory, and incomplete categories), thus ensuring balanced sample sizes. Subsequently, each evaluated model is prompted to classify every task description as clear or unclear. For the mutated task descriptions, we further ask the model to classify them into one of the three issue categories.

Figure \ref{fig:prompt_classification} shows the classification results for seven LLMs, including GPT-4, Qwen, DeepSeek Coder, and CodeLlama, in both their small (6–7B) and large (32–34B) variants.  

The MCC score of the evaluated LLMs in classifying task descriptions as `clear' or `unclear' ranges from -0.06 to 0.47 on HumanEval and from -0.1 to 0.45 on MBPP, with GPT‑4 and both Qwen variants (7B and 32B) achieving the highest values in each dataset. When we ask LLMs to assign each unclear task description to one of the three defined categories (ambiguous, incomplete, and contradictory), GPT4 again leads, recording an MCC of 0.55 on HumanEval and 0.45 on MBPP, followed by Qwen32B at 0.50 and 0.42, respectively; all other models fall between –0.10 and 0.30.

All the LLMs, except Deepseek variants,  classify `Contradictory' descriptions best, then `Incomplete', while `Ambiguous' descriptions are hardest. Deepseek models do not follow a clear order. Overall, `Ambiguous' descriptions remain challenging for almost every model.

Overall, our evaluation reveals that language models like GPT-4 demonstrate only modest capability in identifying unclear task descriptions. Even the best-performing LLMs attain an MCC of approximately 0.50. These models also often fail to pinpoint the issues in a task description; deeply ambiguous cases, where multiple interpretations are equally valid, are the hardest for all architectures tested. Additionally, when prompting the LLMs to explain the issues, they failed badly as in almost all cases what they pointed out was unclear or irrelevant. \\

\noindent
\begin{minipage}{\columnwidth}
\setlength{\fboxsep}{6pt} 
\colorbox{gray!15}{
  \parbox{\dimexpr\columnwidth-2\fboxsep-2\fboxrule}{
    \textbf{RQ1 Results.}
    Code generation models cannot reliably detect problems in task descriptions. This shows that, unlike humans, they cannot natively react to low-quality requirements and would nevertheless attempt to produce a solution.
  }
}
\end{minipage}

\subsection{RQ2 -- Model Robustness}
\label{sec:rq2}

The fact that code generation models cannot detect unclear task descriptions motivate us to study their ability to nevertheless produce the expected code. To assess this robustness, we compare their performance on both HumanEval and MBPP when they are confronted to original, clear task descriptions versus the unclear task descriptions that we generate. 

Table~\ref{tab:llm_eval_humaneval} and Table~\ref{tab:llm_eval_mbpp} show the results on HumanEval and MBPP, respectively. Across both data sets and all models, Pass@1 performance decreases significantly when faced with unclear task descriptions. On average, ambiguous descriptions lead to a 25–30\% reduction in Pass@1 accuracy, while incomplete descriptions cause drops of 20 to 25\%. Contradictory descriptions have the most severe impact, reducing accuracy by up to 40\%. For example, GPT-4 achieves 73.8\% Pass@1 on original HumanEval descriptions, but only 6.7\% on contradictory ones. Similar trends are observed with Qwen-32B and DeepSeek-33B, indicating that even the most capable models are sensitive to quality issues in task descriptions.

A similar trend appears when considering the semantic correctness of the generated code. The Runnable but Incorrect Rate (RIR) are also increasing significantly, e.g., the GPT-4 model increases the incorrect behaviors of the produced code from 24\% in the original descriptions to 54\%, 65\% and 89\% in the documents. cases of Incomplete, Ambiguous and contradictory descriptions. This indicates major issues with the semantic correctness of the generated code even for the cases that the LLMs managed to produce some syntactically valid code. 

Interestingly, the relative impact of quality issues varies between benchmarks \cite{chen2025enhancing}. In HumanEval, ambiguous descriptions tend to degrade performance more than incomplete ones, whereas on MBPP, we observe the opposite. This may be due to the nature of the tasks: HumanEval problems typically require deeper reasoning, making ambiguity more important, while MBPP tasks rely more on explicitly defined operations, where missing details are vital.

\setlength{\tabcolsep}{4pt} 
\begin{table*}[!htbp]
 \vspace{-0.5em}
  \caption{Evaluation of four LLMs on the HumanEval benchmark under four task descriptions conditions (Original, Incomplete, Ambiguous, Contradictory). Metrics reported are Pass@1, Successful Execution Rate (SER), and Runnable but Incorrect Rate (RIR), all expressed as percentages. Red downward arrows (↓) denote the drop in Pass@1 relative to the Original task descriptions.   }
 \vspace{-1.0em}
  \label{tab:llm_eval_humaneval}
  \centering

  \begin{tabular}{|l|ccc|ccc|ccc|ccc|}
    \hline
    \multirow{2}{*}{\textbf{Model}} &
      \multicolumn{3}{c|}{\textbf{Original}} &
      \multicolumn{3}{c|}{\textbf{Incomplete}} &
      \multicolumn{3}{c|}{\textbf{Ambiguous}} &
      \multicolumn{3}{c|}{\textbf{Contradictory}} \\
    \cline{2-13}
      & \textbf{Pass@1} & \textbf{SER} & \textbf{RIR} &
        \textbf{Pass@1} & \textbf{SER} & \textbf{RIR} &
        \textbf{Pass@1} & \textbf{SER} & \textbf{RIR} &
        \textbf{Pass@1} & \textbf{SER} & \textbf{RIR} \\ 
    \hline

    \multicolumn{13}{|c|}{\textbf{Smaller models ( 6-7 B params)}} \\ \hline
    CodeLlama-7B-Instruct-hf  &37.8 & 83.5 & 45.7 & 25.0 \Deltapct{-12.8} &84.8 &	59.8 & 24.4 \Deltapct{-13.4} & 79.9	& 55.5 & 4.3 \Deltapct{-33.5}  & 65.2 & 61.0 \\
    deepseek-coder-6.7B-instruct    &75.6 & 95.7&	20.1 &  53.0 \Deltapct{-22.6} &92.1	& 39.0 & 41.5 \Deltapct{-34.1}	& 86.6	& 45.1 & 8.5  \Deltapct{-67.1} & 80.5 & 72 \\
    Qwen2.5-7B-Instruct  & 79.3 &  96.3	& 17.1  &56.1 \Deltapct{-23.2}  & 94.5 &	38.4 & 54.9  \Deltapct{-24.4} & 92.7	& 37.8 & 4.3 \Deltapct{-75.0} & 75.6 &71.3 \\
    \hline
    \multicolumn{13}{|c|}{\textbf{Larger models 32- 34 B params}} \\ \hline
    GPT-4          &73.8 & 98	 &	24.4 & 45.1 \Deltapct{-28.7} &  98.0 &	53.7 & 34.8  \Deltapct{-39} & 99	&	64.6 & 6.7 \Deltapct{-67.1}  & 95.0 &	89 \\
     CodeLlama-34b-Instruct-hf  & 50.0 & 86.0 & 36.0 & 29.3 \Deltapct{-20.7} & 78.7 & 49.4 & 26.8 \Deltapct{-23.2}  & 79.3 & 52.4 & 4.9 \Deltapct{-45.1} & 70.1 & 65.2 \\
    deepseek-coder-33b-instruct  & 71.3 & 95.7	&24.4 & 48.8 \Deltapct{-22.5}  & 91.5	&	42.7 & 44.5 \Deltapct{-26.8} & 86.0 & 41.5 & 9.1 \Deltapct{-61.6} & 76.8 & 67.7 \\
    Qwen2.5-32B-Instruct       &  86.0 &  97.0	&11.0  & 61.1 \Deltapct{-24.9 } & 92.7 &	31.1 & 56.1   \Deltapct{-29.9 }     & 94.5	&38.4 & 8.5  \Deltapct{-77.5 }  & 81.7&	73.2 \\
  
    \hline
    
  \end{tabular}
\end{table*}

\begin{table*}[!htbp]
  \vspace{-0.5em}
  \caption{ Evaluation results on the MBPP benchmark. }
  \vspace{-1.0em}
  \label{tab:llm_eval_mbpp}
  \centering

  \begin{tabular}{|l|ccc|ccc|ccc|ccc|}
    \hline
    \multirow{2}{*}{\textbf{Model}} &
      \multicolumn{3}{c|}{\textbf{Original}} &
      \multicolumn{3}{c|}{\textbf{Incomplete}} &
      \multicolumn{3}{c|}{\textbf{Ambiguous}} &
      \multicolumn{3}{c|}{\textbf{Contradictory}} \\
    \cline{2-13}
      & \textbf{Pass@1} & \textbf{SER} & \textbf{RIR} &
        \textbf{Pass@1} & \textbf{SER} & \textbf{RIR} &
        \textbf{Pass@1} & \textbf{SER} & \textbf{RIR} &
        \textbf{Pass@1} & \textbf{SER} & \textbf{RIR} \\ 
    \hline

    \multicolumn{13}{|c|}{\textbf{Smaller models ( 6-7 B params)}} \\ \hline
    CodeLlama-7B-Instruct-hf        & 31.9 &   69.7	& 37.8 & 19.6 \Deltapct{- 12.3 }  & 55.5 &	35.9 & 20.3 \Deltapct{-11.6 }  &  55.2	& 	34.9 & 13.6	\Deltapct{ -18.3}  & 67.0	 & 53.5 \\
    
    deepseek-coder-6.7B-instruct    &   43.8 &  77.6& 33.8 & 27.8 \Deltapct{ -16 }  &63.8&	36.4 & 28.0	\Deltapct{-15.8 }  &63.8	&35.7  &  26.7 \Deltapct{ - 17.1}  & 72.0&	45.3 \\
    
    Qwen2.5-7B-Instruct             &46.8 & 78	& 31.2 & 29.4 \Deltapct{ -17.4}  & 62.7	&	33.4 & 30.2 \Deltapct{ -16.6 }  &  62.5 &	32.3 & 14.2 \Deltapct{ -32.6}  & 72&	58.1 \\
    \hline
    \multicolumn{13}{|c|}{\textbf{ Larger models (32- 34 B params)}} \\ \hline
    GPT-4   & 38.6 & 93 &54.6 & 10.4 \Deltapct{ -28.2}  & 98.0 & 88.4 & 15.6 \Deltapct{ -23}  & 95&	80.3 & 	8.4 \Deltapct{ -30.2 }  & 73 &64.9 \\
    
     CodeLlama-34b-Instruct-hf & 37.1 & 72.2 & 35.1  & 21.7 \Deltapct{ -15.4}  & 54.5&	32.9 & 22.6 \Deltapct{ -14.5} & 55 & 32.4 & 17.7 \Deltapct{ -20.0}  & 69.1 & 51.4  \\
     
    deepseek-coder-33b-instruct    & 46.5 & 77.7 & 31.2 &  30.3 \Deltapct{ -16.2 }  & 64.7 & 34.4 & 29.9 \Deltapct{ -16.6}  & 64.2 & 34.3 & 23.7 \Deltapct{ -22.8} & 71.9 & 11 \\
    
    Qwen2.5-32B-Instruct  & 50.3 & 78.3	& 28 & 31.2 \Deltapct{ -19.1}  & 64.4	& 33.2 & 32.8 \Deltapct{ -17.5}& 64.5 & 31.7 & 14.7  \Deltapct{-35.6 } & 73.2 & 58.5  \\
    \hline
  \end{tabular}
\vspace{-0.5em}
\end{table*}
\setlength{\tabcolsep}{6pt}

\begin{figure}[!htbp]
\centering
\vspace{-0.5em}
\includegraphics[width=0.79\linewidth]{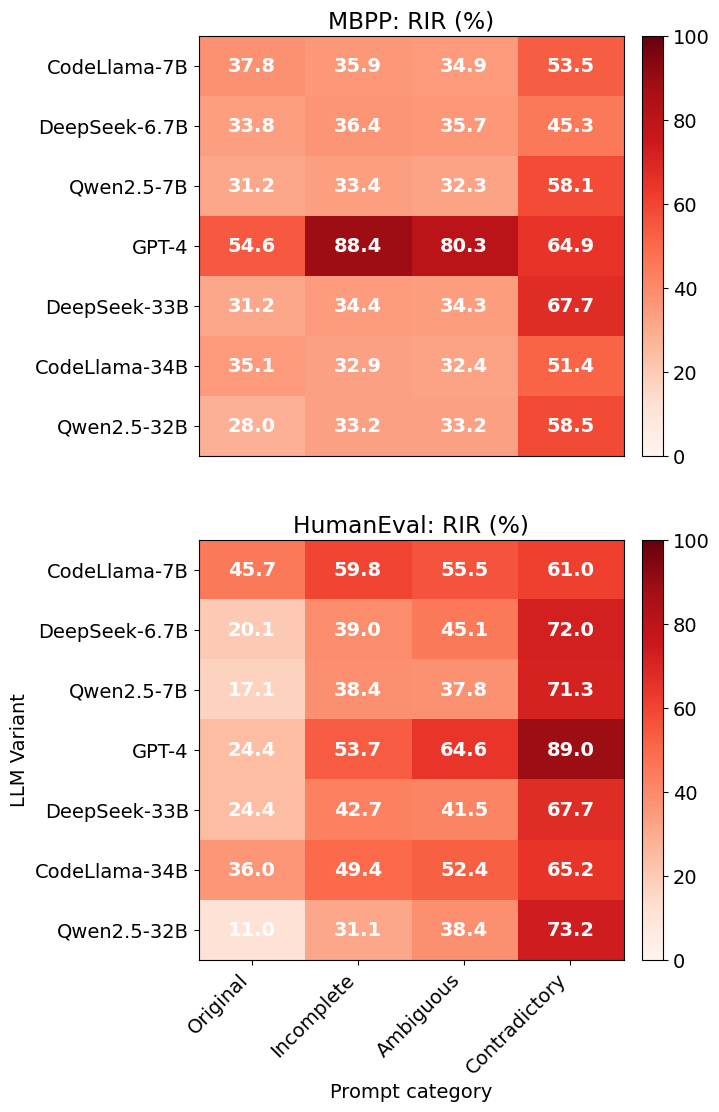} 
\vspace{-1.0em}
\caption{  Runnable but incorrect rates (\%) of seven LLM variants across four task descriptions categories on MBPP (left) and HumanEval (right). Higher values indicate a greater share of executions that run but produce incorrect outputs. }
\label{fig:RIR}
\vspace{-0.5em}
\end{figure}

Although SER remains high across all conditions, Figure~\ref{fig:RIR} shows that RIR increases substantially when task descriptions quality drops \cite{rossiai}. This suggests that while models continue to produce syntactically valid code, a large fraction of the outputs are semantically incorrect. For instance, under contradictory task descriptions, RIR exceeds 80\% for GPT-4 and several other models. 

\begin{figure*}[!htbp]
\centering
\vspace{-0.5em}
\includegraphics[width=0.7\textwidth]{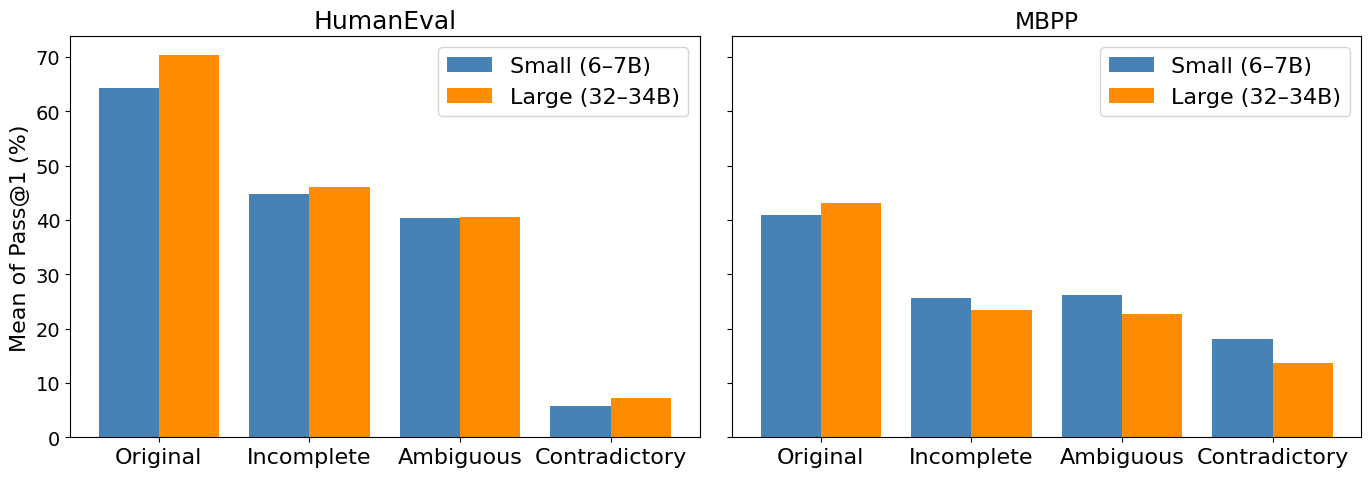} 
\vspace{-1.0em}
\caption{ Mean Pass@1 accuracy (\%) of small (6–7B) versus large (32–34B) model variants on HumanEval (left) and MBPP (right) across four task description categories: Original, Incomplete, Ambiguous, and Contradictory. }
\label{fig:large_small_llms}
\vspace{-0.5em}
\end{figure*}

Figure \ref{fig:large_small_llms} shows that LLM size does affect robustness, though the extent varies across benchmarks. On HumanEval, larger models (32–34B) slightly outperform their smaller counterparts (6–7B) across all descriptions types, particularly under `incomplete' and `contradictory' categories. For example, Qwen-32B achieves notably higher Pass@1 scores than Qwen-7B across the board. However, in MBPP, this trend is less consistent: the performance gap between large and small models is narrower, and in most cases, smaller models perform comparably or even better. This suggests that while increasing model size can improve robustness, particularly on benchmarks like HumanEval that feature more diverse and open-ended tasks, it does not uniformly translate to better performance across all settings. In MBPP, where task descriptions tend to be shorter and more narrowly scoped, the advantages of scale appear more limited.

As shown in Tables \ref{tab:llm_eval_humaneval} \ref{tab:llm_eval_mbpp}, Qwen models with both variants (7B and 32B) are generally the top-performing models, followed by DeepSeek models and GPT4. CodeLlama models consistently underperform, especially under ambiguous and contradictory task descriptions. These differences may stem from variations in pretraining corpora, instruction tuning, or architectural design. 

In summary, the results clearly show that task descriptions' imperfections have a substantial negative impact on LLM performance, even for state of the art models. Larger models exhibit better robustness, but remain vulnerable to ambiguity and contradiction. These observations suggest that future work should focus on improving model interpretability, incorporating description verification mechanisms, and finetuning on noisy or unclear data to better reflect real world development scenarios.\\

\noindent
\begin{minipage}{\columnwidth}
\setlength{\fboxsep}{6pt} 
\colorbox{gray!15}{
  \parbox{\dimexpr\columnwidth-2\fboxsep-2\fboxrule}{
    \textbf{RQ2 Results.} Task description issues lead to a 20–40\% drop in Pass@1 in HumanEval and MBPP benchmarks, most pronounced under contradictory task descriptions, and raise the rate of runnable‑but‑incorrect outputs. While larger models (32–34B) and Qwen variants exhibit comparatively greater resilience, all architectures remain susceptible to ambiguous or incomplete task specifications. Nonetheless, the semantic correctness of the generated code shows a substantial decrease in relation to the quality of task descriptions making the generated code behave incorrectly in 60-90\%  of the cases. 
  }
}
\end{minipage}


\subsection{RQ3 -- Error Type Analysis}
To better understand the nature of model failures under unclear requirements, we analyze not only whether a generated solution runs successfully, but also whether it produces the correct logic. These findings demonstrate that unclear requirements not only degrade task performance but also alter the nature of failure modes \cite{dou2024s}. Although all models maintained a high Successful Execution Rate (SER) across task descriptions conditions, this metric alone is insufficient. As shown in Figure~\ref{fig:RIR}, the Runnable but Incorrect Rate (RIR) increases significantly as description clarity declines. This means models often produce syntactically valid code that compiles and executes, yet fails the functional requirements of the task—suggesting deep semantic misunderstandings. For instance, under contradictory task descriptions, RIR exceeds 80(\%) for GPT-4 and approaches similar levels for other large models. In the case of ambiguous task descriptions, RIR remains high across all LLMs, with some small models such as Qwen-7B reaching over 70(\%). These results reinforce that execution alone is not a reliable proxy for correctness, especially when task descriptions contain subtle or conflicting defects. To further unpack these failures, we analyzed the types of exceptions thrown by incorrect generations using our largest LLMs. We extract 3114 errors and categorize them in Table~\ref{tab:errors}, showing the distribution of common Python exception types across each descriptions category. Our findings reveal that:

\begin{itemize}
    \item \textbf{Incomplete task descriptions:}  are particularly prone to fundamental runtime failures \cite{atiyaa2025exploring} such as NotFoundError (82.4(\%)), TypeError (37.3(\%)), and SyntaxError (41.3(\%)). These reflect structural gaps in the task description, often leading models to hallucinate or mishandle variables, arguments, or input types.
    \item \textbf{Ambiguous task descriptions:} in contrast, tend to generate semantically flawed logic that nonetheless executes. These are marked by elevated AttributeError (31.4(\%)) and KeyError (38.9(\%)), which typically emerge from misinterpreting vague specifications or assuming incorrect object structures.
    \item \textbf{Contradictory task descriptions} result in a mixed profile of errors, with moderate rates of logical faults across most categories, but notably higher NameError and ValueError rates. This suggests models are confused by internal inconsistencies, leading to conflicting variable usage or return semantics.
    \item \textbf{Original task descriptions}, though well formed, still yield a notable share of NameError (29.8(\%)) and ValueError (30.1(\%)). Interestingly, these are less common in ambiguous task descriptions, possibly because detailed instructions constrain the model’s solution path, making small mistakes more impactful, while ambiguous task descriptions allow greater structural flexibility.
\end{itemize}

These findings demonstrate that unclear requirements not only degrade task performance but also alter the nature of failure modes. Incomplete task descriptions are more likely to trigger structural issues such as syntax and type errors; ambiguous task descriptions often result in semantically misaligned outputs; and contradictory task descriptions tend to produce logically inconsistent or invalid solutions. This highlights the importance of advancing LLMs’ ability to reason under unclear instructions, a critical step for safe and effective integration into real-world software development workflows where requirement clarity cannot be guaranteed.\\

\begin{table*}[!htbp]
\centering
\caption{Distribution of error types across description categories. Rows (1–4) denote each category’s share of that error type, with the highest in each column highlighted in bold. The ``Total'' row reports the overall count of errors for each type.}
\label{tab:errors}
\renewcommand{\arraystretch}{1.3}

\begin{tabular}{|c|c|c|c|c|c|c|c|c|c|}
\hline
\textbf{Description Category} &
\makecell{\textbf{Attribute}\\\textbf{Error}} &
\makecell{\textbf{Indentation}\\\textbf{Error}} &
\makecell{\textbf{Index}\\\textbf{Error}} &
\makecell{\textbf{Key}\\\textbf{Error}} &
\makecell{\textbf{NotFound}\\\textbf{Error}} &
\makecell{\textbf{Name}\\\textbf{Error}} &
\makecell{\textbf{Syntax}\\\textbf{Error}} &
\makecell{\textbf{Type}\\\textbf{Error}} &
\makecell{\textbf{Value}\\\textbf{Error}} \\
\hline
Ambiguous     & 31.4\% & 28.6\% & 28.2\% & \textbf{38.9\%} & 17.6\% & 26.7\% & 25.4\% & 29.4\% & 21.4\% \\
Contradictory & 21.6\% & 3.4\%  & 17.9\% & 22.2\%          & 0.0\%  & 22.9\% & 22.2\% & 11.5\% & 20.4\% \\
Incomplete    & \textbf{33.3\%} & \textbf{52.1\%} & \textbf{30.8\%} & 16.7\% & \textbf{82.4\%} & 20.6\% & \textbf{41.3\%} & \textbf{37.3\%} & 28.2\% \\
Original      & 13.7\% & 16.0\% & 23.1\% & 22.2\%          & 0.0\%  & \textbf{29.8\%} & 11.1\% & 21.7\% & \textbf{30.1\%} \\
\hline
\textbf{Total} & \textbf{51} & \textbf{119} & \textbf{39} & \textbf{18} & \textbf{17} & \textbf{131} & \textbf{63} & \textbf{2573} & \textbf{103} \\
\hline
\end{tabular}
\end{table*}

\noindent
\begin{minipage}{\columnwidth}
\setlength{\fboxsep}{6pt} 
\colorbox{gray!15}{
  \parbox{\dimexpr\columnwidth-2\fboxsep-2\fboxrule}{
    \textbf{RQ3 Results.} Unclear task descriptions not only lower overall performance but also shift the nature of model failures. Incomplete descriptions lead to structural errors (e.g., SyntaxError, TypeError), ambiguous descriptions cause semantically incorrect but executable code (e.g., AttributeError, KeyError), and contradictory task descriptions produce logically inconsistent behavior (e.g., NameError, ValueError). These distinct error patterns highlight the need for input-sensitive debugging and adaptive mitigation strategies in LLM-powered development workflows.
  }
}
\end{minipage}

%% file: Threats.tex
\section{Threats to validity }

\textbf{External Validity :} Our findings are derived from Python code generation tasks, which may limit the generalizability of our findings. We performed our experiments on two popular benchmarks \cite{mahmood2022software} (HumanEval and MBPP), focusing on relatively small, self-contained functions. These tasks, while varied in topic and complexity, do not cover other programming languages \cite{wang2024your}, multimodule projects, or domain-specific coding scenarios. We consider that in more complex cases, LLMs will perform worse than in our case since the tasks will be more challenging and will offer more opportunities for task description defects. Moreover, LLM performance will be anyway lower when faced with more complex cases. Additonally, the description flaws we introduced (namely ambiguity, incompleteness, and contradiction) represent three common requirement issues but not the full spectrum of possible task descriptions imperfections (e.g., overly verbose or noisy descriptions). We partly addressed generalizability by confirming that robustness trends held across two distinct benchmarks and observing consistent patterns for multiple model families. We evaluated seven state-of-the-art code models, including GPT-4, DeepSeek Coder, CodeLlama, and Qwen in both small and large variants (6–7B and 32–34B). This diversity of models and tasks increases confidence that our conclusions are not tied to a single architecture or dataset.

\textbf{Internal validity : } The methodology we used for generating flawed task descriptions may constitute a threat to the internal validity. We created ambiguous, incomplete, and contradictory task descriptions by prompting GPT-4 to mutate the original benchmarks. This reliance on GPT-4 could introduce artifacts or systematic biases in the mutated task descriptions. We mitigated this risk via rigorous manual verification; multiple reviewers manually inspected and corrected (where needed) each generated task description to ensure it faithfully represented the intended ambiguity/inconsistency and retained fidelity to the original task. Another potential confounding factor is the influence of model randomness on its outcomes. We used a consistent generation setting (greedy decoding) to limit nondeterminism, but if different sampling parameters were used, the success rates might vary. Finally, some original tasks may have been seen in model training data, inflating baseline performance. As previous work shows, even minor wording changes can cause sharp drops for memorized solutions \cite{wang2022recode}. Our task descriptions mutations might break exact overlaps with training data, as such performance degradation may partly stem from reducing such a familiarity advantage rather than from just the model’s reasoning flaws, an unavoidable side-effect when evaluating robustness to description changes.

%% file: Conclusion.tex
\section{Conclusion }
As large language models continue to power code generation tools in real-world development environments, ensuring their robustness to unclear, user written task descriptions becomes increasingly critical. In this study, we conducted a systematic empirical analysis of LLM performance under flawed task descriptions by focusing on three prevalent issues: ambiguity, incompleteness, and contradiction.

Using controlled mutations of HumanEval and MBPP benchmarks, we showed that even minor task descriptions imperfections can lead to significant drops in Pass@1 accuracy and a rise in semantically incorrect but executable code (60-90\% of the code that compiles is semantically incorrect). Although larger models generally outperform smaller ones in terms of compilation rates, the code they produce is (largely) semantically incorrect code. This menas that scale alone does not guarantee robustness especially under contradictory or vague requirements.

Our analysis further reveals that different description flaws induce distinct failure patterns, from structural issues in incomplete task descriptions to semantic misalignment in ambiguous ones. These findings underscore the need for LLMs that are not only accurate but also resilient to the different kinds of unclear requirements frequently encountered in real world development \cite{ritu2025enhancing}.
Future research should investigate training strategies and diagnostic techniques that enable models to better interpret, detect, and recover from vague or under-specified task descriptions.

%% file: dataset_samples.tex
\begin{table*}[tb]
\caption{Examples of mutated task descriptions for both HumanEval and MBPP datasets across the three description defect categories: Incomplete, Ambiguous, and Contradictory. }
\label{tab:mutation_examples}
\centering
\small 
\begin{tabular}{|l|
>{\raggedright\arraybackslash}p{3.5cm}|
>{\raggedright\arraybackslash}p{2.5cm}|
>{\raggedright\arraybackslash}p{4cm}|
>{\raggedright\arraybackslash}p{4cm}|}
\hline
ID & Original Description & Incomplete & Ambiguous & Contradictory \\
\hline
MBPP/2 &
\begin{minipage}[t]{\linewidth}
\raggedright
Write a function to find the similar elements from the given two tuple lists.
\end{minipage}
&
\begin{minipage}[t]{\linewidth}
\raggedright
Write a function to find similar elements shared between lists.
\end{minipage}
&
\begin{minipage}[t]{\linewidth}
\raggedright
Write a function to compare groups and find equal items. 
\end{minipage}
&
\begin{minipage}[t]{\linewidth}
\raggedright
Write a function to find similar elements from the given two tuple lists, but only include elements not in both. 
\end{minipage}

\\
\hline
MBPP/6 &
\begin{minipage}[t]{\linewidth}
\raggedright
Write a python function to check whether the two numbers differ at one bit position only or not.
\end{minipage}
&
\begin{minipage}[t]{\linewidth}
\raggedright
Write a function to check if two numbers differ.
\end{minipage}
&
\begin{minipage}[t]{\linewidth}
\raggedright
Write a function that checks if two numerical values are distinguishable through a minimal binary alteration. 
\end{minipage}
&
\begin{minipage}[t]{\linewidth}
\raggedright
Write a python function to check whether two numbers differ at one bit position or two.
\end{minipage}

\\
\hline
MBPP/10 &
\begin{minipage}[t]{\linewidth}
\raggedright
Write a function to get the n smallest items from a dataset.
\end{minipage}
&
\begin{minipage}[t]{\linewidth}
\raggedright
Write a function to get items from a dataset.
\end{minipage}
&
\begin{minipage}[t]{\linewidth}
\raggedright
 Write a function to select the smallest of a data group. 
\end{minipage}
&
\begin{minipage}[t]{\linewidth}
\raggedright
Write a function to get the n smallest items from a dataset, but return one item only.
\end{minipage}
\\
\hline
MBPP/22 &
\begin{minipage}[t]{\linewidth}
\raggedright
Write a function to find the first duplicate element in a given array of integers.
\end{minipage}
&
\begin{minipage}[t]{\linewidth}
\raggedright
Write a function to find the first duplicate in an array.
\end{minipage}
&
\begin{minipage}[t]{\linewidth}
\raggedright
Write a function to identify first repeating items in a group of numbers. 
\end{minipage}
&
\begin{minipage}[t]{\linewidth}
\raggedright
Write a function to find the first duplicate element in a given array,return last one.
\end{minipage}
\\
\hline
MBPP/34 &
\begin{minipage}[t]{\linewidth}
\raggedright
Write a python function to find the missing number in a sorted array.
\end{minipage}
&
\begin{minipage}[t]{\linewidth}
\raggedright
Write a function to find a missing value in a list.
\end{minipage}
&
\begin{minipage}[t]{\linewidth}
\raggedright
Write a function to look for gaps in a sequence.
\end{minipage}
&
\begin{minipage}[t]{\linewidth}
\raggedright
Write a python function to find the missing number in a sorted array, but always return the first number.
\end{minipage}
\\
\hline
HumanEval/13 & def greatest\_common\_divisor(a: int, b: int) -\textgreater{} int: Return the greatest common divisor of two integers a and b \textgreater{}\textgreater{}\textgreater{} greatest\_common\_divisor(3, 5) 1 \textgreater{}\textgreater{}\textgreater{} greatest\_common\_divisor(25, 15) 5
&
\begin{minipage}[t]{\linewidth}
\raggedright
Write a program that calculate the common divisor  with two values
\end{minipage}
&
\begin{minipage}[t]{\linewidth}
\raggedright
Write a Python instruction that finds the biggest whole number that implie divisibility exactly into two given amounts. 
\end{minipage}
&def greatest\_common\_divisor(a: str, b: str) -\textgreater{} str:  Return a greatest common divisor of two integers a and b
\\
\hline
HumanEval/16 & def count\_distinct\_characters(string: str) -\textgreater{} int: """ Given a string, find out how many distinct characters (regardless of case) does it consist of \textgreater{}\textgreater{}\textgreater{} count\_distinct\_characters
('xyzXYZ') 3 \textgreater{}\textgreater{}\textgreater{} count\_distinct\_characters
('Jerry') 4 """
&
def count\_distinct\_charact
ers(string):""" find out how many distinct characters does it consist of. """

& Come up with a way to figure out how many different 
letters show up in a word, treating big and small 
versions of the same letter as the same thing.
 & def count\_distinct\_characters(string: str, ignore\_case: bool) -\textgreater{} str: """ Given a string, find out how many distinct characters it consists. please return the number of similar characters. """
\\
\hline

HumanEval/23 & 
def strlen(string: str) -> int:
    """ Return length of given string
    >>> strlen('')
    0
    >>> strlen('abc')
    3
    """
&
def strlen:\newline
\hspace*{1em}""" Return length of string """
&
Create a way to determine how much content is in the given piece of text.
&
def strlen(string: str) -> str:\newline
\hspace*{1em}""" Return length of given string in string format without using the len function.\newline
\hspace*{1em}>>> strlen(11)\newline
\hspace*{1em}0\newline
\hspace*{1em}"""
\\
\hline
HumanEval/27 & def flip\_case(string: str) -\textgreater{} str: """ For a given string, flip lowercase characters to uppercase and uppercase to lowercase. \textgreater{}\textgreater{}\textgreater{} flip\_case('Hello') 'hELLO' """

&
\begin{minipage}[t]{\linewidth}
\raggedright
 For a given input, flip lowercase characters to uppercase and uppercase to lowercase.
\end{minipage}
&
\begin{minipage}[t]{\linewidth}
\raggedright
Given a string, transform each letter so that those that normally stand tall now crouch, and those that usually crouch stand tall.
\end{minipage}
&Take a piece of text and transform it so that all letters that are normally written in a smaller form become capitalized, while also ensuring that no uppercase letters are altered in any way.
\\
\hline
HumanEval/28 & from typing import List def concatenate(strings: List[str]) -\textgreater{} str: """ Concatenate list of strings into a single string \textgreater{}\textgreater{}\textgreater{} concatenate([]) '' \textgreater{}\textgreater{}\textgreater{} concatenate(['a', 'b', 'c']) 'abc' """
&
\begin{minipage}[t]{\linewidth}
\raggedright
Concatenate list of strings into a single string.
\end{minipage}
&
\begin{minipage}[t]{\linewidth}
\raggedright
Given a bunch of pieces of text, figure out how to make them into one continuous bit.
\end{minipage}
&from typing import List def concatenate(strings: List[str]) -\textgreater{} int: """ Concatenate list of strings into a single string but return an integer. The function should also work for lists of integers. """
\\
\hline
\end{tabular}
\end{table*}